\documentclass[usenatbib]{mn2e}
\usepackage{natbibmnfix,graphicx,times}

\newcommand{\cmden}{\mbox{ cm$^{-3}$}}
\newcommand{\colden}{\mbox{ cm$^{-2}$}}

\newcommand{\kel}{\mbox{ K}}

\newcommand{\Mpc}{\mbox{ Mpc}}
\newcommand{\kpc}{\mbox{ kpc}}

\newcommand{\Zsun}{\mbox{ Z$_\odot$}}
\newcommand{\secinv}{\mbox{ s$^{-1}$}}
\newcommand{\hunits}{\mbox{ km s$^{-1}$ Mpc$^{-1}$}}
\newcommand{\kms}{\mbox{ km s$^{-1}$}}
\newcommand{\bq}{\begin{equation}}
\newcommand{\eq}{\end{equation}}
\newcommand{\bqa}{\begin{eqnarray}}
\newcommand{\eqa}{\end{eqnarray}}

\newcommand{\osix}{O{\sc VI} }
\newcommand{\oseven}{O{\sc VII} }
\newcommand{\oeight}{O{\sc VIII} }
\newcommand{\osixns}{O{\sc VI}}
\newcommand{\osevenns}{O{\sc VII}}
\newcommand{\oeightns}{O{\sc VIII}}
\newcommand{\Nosix}{N_{\rm OVI}}
\newcommand{\Noseven}{N_{\rm OVII}}
\newcommand{\Noeight}{N_{\rm OVIII}}
\newcommand{\nosix}{n_{\rm OVI}}
\newcommand{\noseven}{n_{\rm OVII}}
\newcommand{\noeight}{n_{\rm OVIII}}
\newcommand{\Nion}{N_{\rm ion}}
\newcommand{\fosix}{f_{\rm OVI}}
\newcommand{\foseven}{f_{\rm OVII}}
\newcommand{\foeight}{f_{\rm OVIII}}

\newcommand{\deriv}{{\rm d}}

\title[Oxygen Absorbers in the IGM]{Highly-Ionized Oxygen Absorbers in the Intergalactic Medium}

\author[S.~Furlanetto, L.~A. Phillips, \& M. Kamionkowski]{S.~R. Furlanetto,\thanks{Email:sfurlane@tapir.caltech.edu} L.~A. Phillips, and M. Kamionkowski \\
California Institute of Technology, Mail Code 130-33, Pasadena, CA 91125, USA}

\begin{document}

\maketitle

\begin{abstract}
Recent ultraviolet and X-ray observations of intergalactic \osix
and \oseven absorption systems along lines of sight to bright
quasars have opened a new window onto the ``warm-hot
intergalactic medium'' (WHIM).  These systems appear to provide
a significant reservoir for baryons in the local universe, and
comparison to cosmological simulations suggests that their
abundance roughly matches theoretical predictions.  Here we use
analytic arguments to elucidate the physical properties of the
absorbers and their role in structure formation.  We first show
that if the absorbers result from structure-formation shocks,
the observed column densities naturally follow from
postshock-cooling models, if we include fast-cooling shocks as well as those that cannot cool within a  Hubble time.  In this case, the known \osix absorbers should show stronger \oseven
absorption than expected from collisional-ionization equilibrium
(and much more than expected for photoionized systems).  We then
argue that higher-temperature shocks will be spatially
associated with more massive virialized objects even well
outside the virial radius.  Thus the different oxygen ions will
trace different structures; \oseven absorbers are the most
common because that ion dominates over a wide temperature range
(corresponding to a large range in halo mass).  If each dark-matter halo is surrounded by a network
of shocks with total cross section a few times the size of the
virialized systems, then we can reproduce the observed number densities
of absorbers with plausible parameters.  A simple comparison with
simulations shows that these assumptions are reasonable,
although the actual distribution of shocked gas is too complex
for analytic models to describe fully.  Our models suggest that
these absorbers cannot be explained as a single-temperature phase.
\end{abstract}


\section{Introduction}
\label{intro}

The cosmic-web paradigm has had a great deal of success in explaining the distribution of matter in the intergalactic medium (IGM) at moderate redshifts $z \sim 3$, especially in the Ly$\alpha$ forest (e.g., \citealt{rauch98}).  In this picture, dark matter collapses through gravitational instability into sheets and filaments before accreting onto bound structures such as galaxies, groups, and clusters (which appear at the intersections of filaments).  The baryons also accrete onto these structures, but (in addition to gravitational forces) they are subject to fluid effects such as shock heating.  At $z \sim 3$, the infall velocities are modest relative to the ambient sound speed and shocks are relatively unimportant except in collapsed objects.  However, at the present day, shocks have heated most of the filamentary gas to $T \ga 10^5 \kel$ \citep{cen99}; this gas is now known as the ``warm-hot IGM'' (WHIM).  Simulations predict that this phase contains a substantial fraction of the baryons \citep{cen99,dave01-whim}.  The characteristic temperature $T \sim 10^6 \kel$ corresponds to the postshock temperature that has collapsed over the nonlinear mass scale in one Hubble time.

Unfortunately, most of the WHIM lies at a modest density ($\sim 10$--$100$ times the cosmic mean) and in a temperature regime with few observational signatures; the hydrogen column densities are small because of collisional ionization and the gas is not hot enough to emit substantial bremsstrahlung radiation.  As such, the WHIM has thus far only been detected through absorption of highly-ionized metals (specifically \osixns, \osevenns, and possibly \oeightns).  Space-based ultraviolet spectrographs (including STIS on the \emph{Hubble Space Telescope} and the \emph{Far Ultraviolet Spectroscopic Explorer}) have allowed relatively straightforward detections of the \osix $\lambda \lambda 1032,1038$ doublet along lines of sight to nearby bright quasars \citep{tripp00,tripp00b,oegerle00,sembach01,savage02,shull03,richter04,prochaska04,tumlinson05,danforth05}.  The inferred number density of absorption systems is large ($\deriv n/\deriv z \sim 15$), suggesting that these systems constitute a significant baryon reservoir.  However, in collisional-ionization equilibrium (CIE), \osix only exists near the lower end of the WHIM temperature range.  According to the simulations, most of the gas can only be probed through \oseven or \oeight absorption, both of which require X-ray studies.  The current generation of instruments does not have the sensitivity to measure the weak absorption expected from the IGM, and there are relatively few firm detections.  A number of groups have found \oseven and \oeight absorption at $z=0$ that may be associated with the Local Group \citep{nicastro02,rasmussen03,fang03}.  \citet{nicastro04} recently discovered two \oseven absorbers along the line of sight to the blazar Mkn 421 during an outburst.  Other detections have relatively low significance or lack confirmation from different instruments (e.g., \citealt{fang02b}).  Nevertheless, the Mkn 421 line of sight suggests an even higher number density than \osix ($\deriv n/\deriv z \sim 60$), albeit with large statistical uncertainties.

Comparison to lines of sight extracted from simulations suggests that these number densities are comparable to theoretical expectations \citep{hellsten98,cen01,fang01,fang02,chen03}.  Detailed quantitative tests are difficult because of the large uncertainties on both the observational and theoretical sides (with the latter depending primarily on the unknown distribution of metals in the IGM), but the overall consistency is a reassuring confirmation of the cosmic-web paradigm.  However, the existing studies have not addressed many questions about their physical origin or characteristic environment.  As a result, we lack a clear picture of their specific role in the structure-formation process.  For example, do they probe outlying filamentary gas, gas more closely associated with galaxies, or something entirely different?  Do the \osix and \oseven absorbers probe distinct environments, and do the two ions have different physical origins?  What sets the observed column densities?

In this paper, we provide an analytic framework to answer these questions.  This is clearly a difficult proposition:  the WHIM is nonlinear and inherently asymmetric, so it is difficult to model analytically.  Nevertheless, models describing the gross properties of the WHIM do exist \citep{nath01,valageas02,furl04-sh}, and it is interesting to consider what these simple arguments reveal about the absorbers.  Moreover, if shocks are ultimately responsible for heating gas to the WHIM temperatures, simple models of postshock cooling can help to explain the properties of the absorbers \citep{heckman02}.  We will connect these two approaches in order to identify the key properties of the absorbers and how they may relate to collapsed objects.

The plan of this paper is as follows.  In \S \ref{ncol}, we show how postshock cooling predicts characteristic column densities for the absorbers.  We then construct a simple analytic model to predict the abundance of IGM oxygen absorbers in \S \ref{abundance}.  We highlight some of the strengths and weaknesses of this model by comparing to cosmological simulations in \S \ref{sims}.  In \S \ref{photo}, we show that comparing the \osix and \oseven columns can robustly distinguish shock-heated and photoionized \osix absorbers.  Finally, we conclude in \S \ref{disc}.

We will primarily be concerned with \osixns, \osevenns, and \oeight in this paper.  For reference, we list details of their strongest transitions here.  \osix has a UV doublet at $\lambda=1032,\, 1038$ \AA; the corresponding oscillator strengths are $f_{\rm osc}^{\rm OVI}=0.133,\,0.0660$.  The most prominent \oseven line  is a He$\alpha$ transition with $\lambda=21.602$ \AA \ and $f_{\rm osc}^{\rm OVII}=0.6945$, while that for \oeight is a Ly$\alpha$ transition with $\lambda=18.97$ \AA \ and $f_{\rm osc}^{\rm OVIII}=0.416$.  To avoid any assumptions about the velocity widths of each line, we will present all of our results in terms of column density $\Nion$.  Observations are instead often presented in terms of equivalent width $W$ (which is the fundamental observable).  In the unsaturated limit, the two quantities are related (in cgs units) via $W=8.85 \times 10^{-13} \Nion f_{\rm osc} \lambda^2$ (e.g., \citealt{spitzer78}).  (For \osixns, $W$ is usually, though not universally, quoted for the stronger of the two doublet transitions.)

Throughout our discussion we assume a cosmology with $\Omega_m=0.3$, $\Omega_\Lambda=0.7$, $\Omega_b=0.046$, $H_0=100 h \hunits$ (with $h=0.7$), $n=1$, and $\sigma_8=0.9$, consistent with the most recent measurements \citep{spergel03}.

\section{The Characteristic Column Densities}
\label{ncol}

We will begin with the simple question of whether shock heating can account for the observed column densities of \osix absorbers.  A number of recent observations have found a large number of such systems in the local universe, with typical column densities in the range $\Nosix \approx 2$--$20 \times 10^{13} \colden$ (e.g., \citealt{tripp00,sembach01,richter04,prochaska04}).  While the lower limit may be an observational selection effect (but see \citealt{danforth05}), the rapid decline in abundance as the column density approaches $10^{14} \colden$ suggests that this is a characteristic maximum strength.

\citet{heckman02} provide a simple argument for why we should expect shocks to produce columns in this range.  Consider a parcel of gas that is shocked to a temperature $T_s$.  It will flow away from the
shock with a velocity $v_g$ (fixed by the jump conditions).  After a time $t_{\rm cool} = 3 k T_s/n_s \Lambda$ it will have radiated its thermal energy and no longer be highly ionized; here $n_s$ is the postshock gas density and $\Lambda$ its normalized cooling rate in ergs cm$^3$ s$^{-1}$.  If we view
the system along a plane perpendicular to the shock, and \emph{if} the gas cools fully, we therefore expect to see highly-ionized gas along a path length $l_{\rm cool} \approx v_g t_{\rm cool}$ with a
corresponding column density $N \approx n_s l_{\rm cool}$.  Thus
\bq
\Nosix = \fosix \ Z \ [{\rm O/H}]_\odot \ n_s \ v_g \ t_{\rm cool}
\label{eq:no6heck}
\eq
where $[{\rm O/H}]_\odot=10^{-3.07}$ is the oxygen abundance by number in gas with solar metallicity (using the abundances of \citealt{sutherland93}), $Z$ is the metallicity in solar units, and $\fosix$ is the fraction of oxygen atoms in the fifth ionized state.  We note that the postshock speed is directly related to the postshock temperature via
\bq
v_g = \left( \frac{k T_s}{3 \mu m_p } \right)^{1/2} \approx 50 \left( \frac{T_s}{10^6 \kel} \right)^{1/2} \kms,
\label{eq:vg}
\eq
where we have assumed that the shock is strong (so that $v_g = v_s/4$, with $v_s$ the speed of the shock) and $\mu m_p$ is the mean molecular mass.  \citet{heckman02} emphasized that for moderately enriched gas, $\Lambda \propto Z$ (e.g., \citealt{sutherland93}).  In this fully-cooled case, we therefore expect the column density to be \emph{independent} of the metallicity and physical density.  

We note that there is some tentative observational evidence for such a picture.  \citet{shull03} and \citet{tumlinson05} each found pairs of \osix absorbers displaced by $v_s \ga 200 \kms$ from associated HI and/or low-ionization state metal absorbers.  They suggested that the velocity difference may be characteristic of a shock.

\begin{figure}
\begin{center}
\resizebox{8cm}{!}{\includegraphics{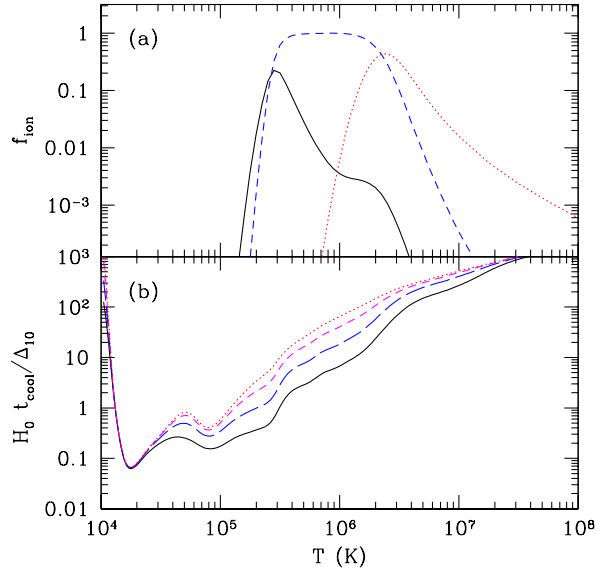}}\\%
\end{center}
\caption{\emph{(a)}: Fractional abundance of \osixns, \osevenns, and \oeight (solid, dashed, and dotted curves, respectively) if we assume CIE.  \emph{(b)}: Ratio of the cooling time to the Hubble time, as a function of gas temperature.  The solid, long-dashed, short-dashed, and dotted curves assume $Z=0.3,\,0.1,\,0.03$, and $0.01 \Zsun$, respectively.  All curves assume that the gas element has density ten times the mean (i.e., $\Delta \equiv 10 \Delta_{10}$, with $\Delta_{10}=1$). }
\label{fig:ionfrac}
\end{figure}

Of course, equation (\ref{eq:no6heck}) is only approximate
because both $t_{\rm cool}$ and $\fosix$ depend on temperature.
To properly compute the column density, we need to trace the
(non-equilibrium) ionic abundances throughout the cooling (e.g.,
\citealt{dopita96}).  Fortunately, we can estimate the result
reasonably well because $\fosix$ peaks sharply around a
characteristic temperature, as we show in
Figure~\ref{fig:ionfrac}\emph{a}.  We have computed the
ionization fractions $f_{\rm ion}$ of the highly-ionized states
of oxygen using Cloudy (version 94, \citealt{cloudy}) for gas in
CIE.  We see that \osix is extremely sensitive to the gas
temperature; $\fosix \sim 0.3$ at $T \approx 10^{5.45} \kel$ but
drops steeply away from this value.  The ionization fraction for
\osevenns, on the other hand, rises rapidly to $\foseven \sim 1$
in the same temperature range for which \osix rises, but it
remains large for all $10^{5.5} \kel < T < 10^{6.5} \kel$.  This
is a simple consequence of the lone $n=2$ electron in \osixns,
which is much easier to strip than the $n=1$ electrons in
\osevenns: the ionization potentials of \osix and \oseven are
$138.1$ and $739.3$ eV, respectively.  The plateau in $\fosix$
at $T\sim 10^6 \kel$ and $\fosix \sim 0.003$ occurs for the same reason.  Clearly, \osix will
come from gas in a narrow temperature range.  Figure
\ref{fig:ionfrac}\emph{b} shows the ratio of the cooling time to
$H_0^{-1}$, also computed with Cloudy,\footnote{The CIE cooling
models of \citet{sutherland93} are often used for this purpose.
However, those tables modify the [O/Fe] ratio as a function of
metallicity; because oxygen is particularly important for our
purposes, we have chosen to keep its abundance fixed relative to
other metals.  For consistency with previous work, we use the
solar-abundance ratios of \citet{sutherland93}, with the
exception of helium (which we set to its primordial value for a
mass fraction $Y=0.24$).} for a range of metallicities.  In the
plot we assume that $\Delta = 10$, where $\Delta = n/\bar{n}$ is
the gas density relative to the cosmic mean; of course $t_{\rm
cool} \propto \Delta^{-1}$.   The cooling time drops rapidly as
we approach $T \sim 10^{5.45} \kel$ from above (partly because of the
strong \osix $\lambda \lambda 1032,\, 1038$ doublet) and then
decreases further as \osix recombines.  \citet{heckman02}
pointed out that this allows us to substitute the values of
$\fosix$ and $t_{\rm cool}$ at the \osix peak in equation
(\ref{eq:no6heck}), yielding
\bq
\Nosix \approx 10^{14} \left( \frac{v_g}{50 \kms} \right) \colden 
\label{eq:no6heck-num}
\eq
for enriched gas; note that the result is independent of metallicity and physical density.  The column density depends \emph{only} on shock velocity ($\Nosix \propto T_s^{1/2}$) in this regime.  However, comparison to the non-equilibrium postshock-cooling models of \citet{dopita96} suggests that our prescription underestimates $\Nosix$ by about a factor of two for most temperatures.  
This is particularly important for our low-density IGM absorbers:  the recombination times for the ions and environments of interest can be comparable to (or even exceed) the age of the universe.  For example, at $\Delta=100$ and $T=10^6$--$10^7 \kel$, we find from Cloudy that (for these three ions) the typical radiative recombination times are, in most cases, only a few times smaller than the Hubble time.  We therefore suggest that detailed shock models will be required for careful quantitative fits to the observations, especially if some shocks occur at lower densities.

The key assumption of this prescription is that the gas can cool past $T \approx 10^{5.45} \kel$.  While that is certainly true for most of the systems \citet{heckman02} considered (such as \osix in galactic winds), it is only marginally satisfied by IGM absorbers.  Even at the \osix peak, $t_{\rm cool} > H_0^{-1}$ for $Z \la 0.05 \Zsun$, and, because metal lines dominate the cooling at $10^{5} \kel \la T \la 10^{6.5} \kel$, it increases rapidly above $T \sim 10^{5.5} \kel$ as more and more atoms are ionized.  Thus, tenuous, metal-poor gas and gas shocked to high temperatures will \emph{not} be able to cool through the \osix peak, and the \citet{heckman02} argument actually only provides an upper limit to column densities in the IGM.  To account for this possibility, we must limit the time over which gas cools to be smaller than $H_0^{-1}$.  We let $T_{\rm min} \equiv T_s [1 - (H_0 t_{\rm cool})^{-1}]$ be the (approximate) temperature to which the gas cools, where the cooling time is evaluated at $T_s$ and $T_{\rm min}=0$ if $t_{\rm cool}<H_0^{-1}$.  We then take $\fosix$ to be equal to its maximum value in the range $(T_{\rm min},T_s)$.  For equation (\ref{eq:no6heck}), we evaluate the cooling time at this effective temperature.  If it exceeds the age of the universe, we instead substitute $H_0^{-1}$.  We note that the solution is more uncertain when $t_{\rm cool} > H_0^{-1}$, because the density and metallicity of the gas do affect the total column density.

\begin{figure}
\begin{center}
\resizebox{8cm}{!}{\includegraphics{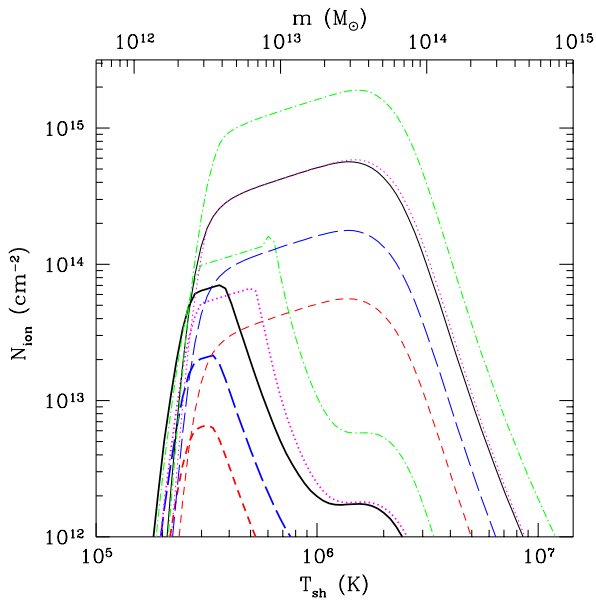}}\\%
\end{center}
\caption{Column density of cooling material corresponding to a given shock temperature.  The lower set of thick lines are for \osixns, while the upper thin lines are for \osevenns.  Solid, long-dashed, and short-dashed curves have $\Delta=10$ with $Z = 10^{-1},\,10^{-1.5}$, and $10^{-2} \Zsun$, respectively.  Dot-dashed and dotted curves have $\Delta=100$ with $Z = 10^{-1.5}$ and $10^{-2} \Zsun$.  The top axis shows the halo mass we associate with each infall shock (see \S \ref{infall}).}
\label{fig:nm}
\end{figure}

We illustrate these features in Figure~\ref{fig:nm}.  The thick curves show the column density of \osix as a function of shock temperature for a range of absorber metallicities and densities.  We emphasize that the maximum $\Nosix$ is nearly constant when $t_{\rm cool} < H_0^{-1}$.  However, for small density or metallicity and for large $T_s$, $\Nosix \propto \Delta Z$.  Thus if \osix absorbers correspond to shocked regions, we expect a characteristic maximum $\Nosix \approx 10^{14} \colden$.  This is only approximate for (at least) two reasons.  First, the viewing angle relative to the shock plane will smear out the sharp maximum.  Second, as described above, detailed non-equilibrium calculations show that the column densities may be somewhat larger than we predict (by a factor of $\sim 2$; \citealt{dopita96}).  The crucial point is that the distribution of column densities should depend on the distribution of shock temperatures in the IGM.  Its maximum is relatively robust to the choice of density and metallicity, but the abundance of weak absorbers can be quite sensitive to those values.

Of course, this postshock-cooling model does not just apply to \osixns.  As \citet{heckman02} pointed out, it also predicts that a range of ionization states should appear and disappear as the gas cools.  Here we will focus on \oseven and \oeightns, whose detection with X-ray telescopes has received a good
deal of attention. Naively comparing $\fosix$ and $\foseven$ in Figure \ref{fig:ionfrac}\emph{a} implies that \emph{every} \osix absorber associated with a shock must also have a large $\Noseven$ (although the converse is not true).  We can use the same procedure as in equation (\ref{eq:no6heck}) to estimate the column density as a function of temperature.  The prescription is actually somewhat simpler because $\foseven$ is essentially flat.  On the other hand, $\foeight$ is large only if $T_s \ga 10^{6.5} \kel$, where $t_{\rm cool} \gg H_0^{-1}$.  Thus we do not expect \osix absorbers to host strong associated \oeight absorption, and vice versa.  Note that this is one important difference between our (time-limited) model and the fully-cooled predictions of \citet{heckman02}.  Of course, \oseven and \oeight can coexist in the same system.

The thin curves in Figure~\ref{fig:nm} show the \oseven column densities corresponding to each shock temperature.  We see that \oseven absorption should be significant over a much larger range of temperatures than \osixns, but because the cooling times are long, the maximum $\Noseven$ is much more sensitive to the gas density and metallicity.  We therefore cannot robustly predict a maximum $\Noseven$.  On the other hand, note that $\Noseven$ is almost always several times larger than $\Nosix$, even near the \osix peak.  

The postshock-cooling model also predicts physical sizes corresponding to each column: $\ell \approx v_g t_{\rm cool} \approx 500 \, (v_g/50 {\rm \ km \ s}^{-1}) \, (H_0t_{\rm cool})$ kpc.  (Note that the last factor cannot exceed unity in our time-limited picture.)  It is possible that geometric effects (such as the filamentary structure of the cosmic web) could provide more stringent constraints on the sizes (and hence column densities) of the absorbers.  Also, note that when the cooling time is long, the density profile of the gas behind the shock can affect the column density (i.e., the assumption of a constant density medium may not be appropriate).

\section{The Abundance of Oxygen Absorbers}
\label{abundance}

\begin{figure}
\begin{center}
\resizebox{8cm}{!}{\includegraphics{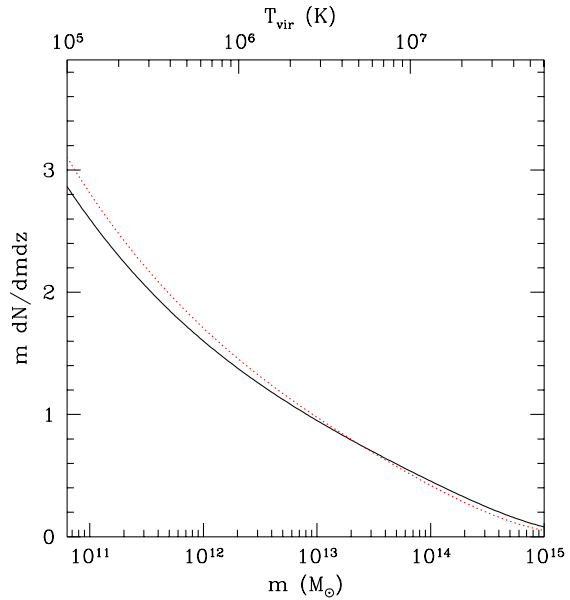}}\\%
\end{center}
\caption{Number of virialized objects intersected per unit redshift as a function of halo mass.  The top axis shows the corresponding virial temperature.  The solid and dotted curves are for $z=0$ and $z=0.2$, respectively. }
\label{fig:cs}
\end{figure}

If the WHIM gas and the oxygen absorbers are a result of structure formation, it is natural to associate them with virialized dark-matter halos.  We will suppose that each halo has a cross-section to host an
absorber that is proportional to the square of its virial radius
$r_{\rm vir}$, which we will define as in \citet{barkana01}.  We
can motivate such a scenario by noting that each halo is
surrounded by a virial shock that heats accreting material to
the virial temperature $T_{\rm vir}$.  Massive galaxy groups and
clusters thus have a hot gas reservoir within $r_{\rm vir}$.
With this picture, we can compute the differential number
$\deriv N/\deriv z$ of absorbers per unit redshift via
\bq
\frac{\deriv N}{\deriv z} = \int \deriv m \,
n(m) \, \pi r_{\rm vir}^2 \frac{\deriv r}{\deriv z},
\label{eq:cs}
\eq
where all quantities are in comoving units.  Here $n(m)$ is the \citet{sheth99} halo mass function.  Figure~\ref{fig:cs} shows the integrand (weighted by mass $m$).  The crucial result is that the
number of systems intersected is small, of order a few per unit redshift.  Thus, to reproduce the observed number density of \osix and \oseven absorbers, we must have absorbers in the IGM outside of virial objects; however, the virialized systems could still account for strong absorbers, so we cannot neglect them entirely.

Moreover, we do not expect every halo to host an absorber of a given type.  As the depth of the potential well increases, the virial shock becomes stronger and the gas temperature increases.  Thus higher
ionization states trace gas associated with larger objects, and we require some prescription to associate dark-matter halos with ion columns.  We will now consider two simple possibilities:  virial shocks and infall shocks.

\subsection{Virialized Halos}
\label{virial}

Once gas falls through the virial shock, it is heated to $T_{\rm vir}$ and begins to cool radiatively. If $t_{\rm cool}$ is small, it accretes onto the central galaxy, passing through a range of ionization states along the way.  If $t_{\rm cool}$ is large, the gas remains in a diffuse halo \citep{white91}.  Virial shocks thus provide an obvious source for hot absorbing gas.\footnote{We note that both simulations and analytic arguments suggest that virial shocks do not form in many systems \citep{birnboim03,keres04}.  However, for the halo masses and redshifts considered here, the ``hot'' accretion phase dominates and virial shocks should be omnipresent.}  The relevant halo masses can be read off from the top
axis of Figure~\ref{fig:cs}, which shows $T_{\rm vir}$ as a function of halo mass: we see that massive galaxies and groups correspond to the temperature range of interest.

\begin{figure*}
\begin{center}
\resizebox{8cm}{!}{\includegraphics{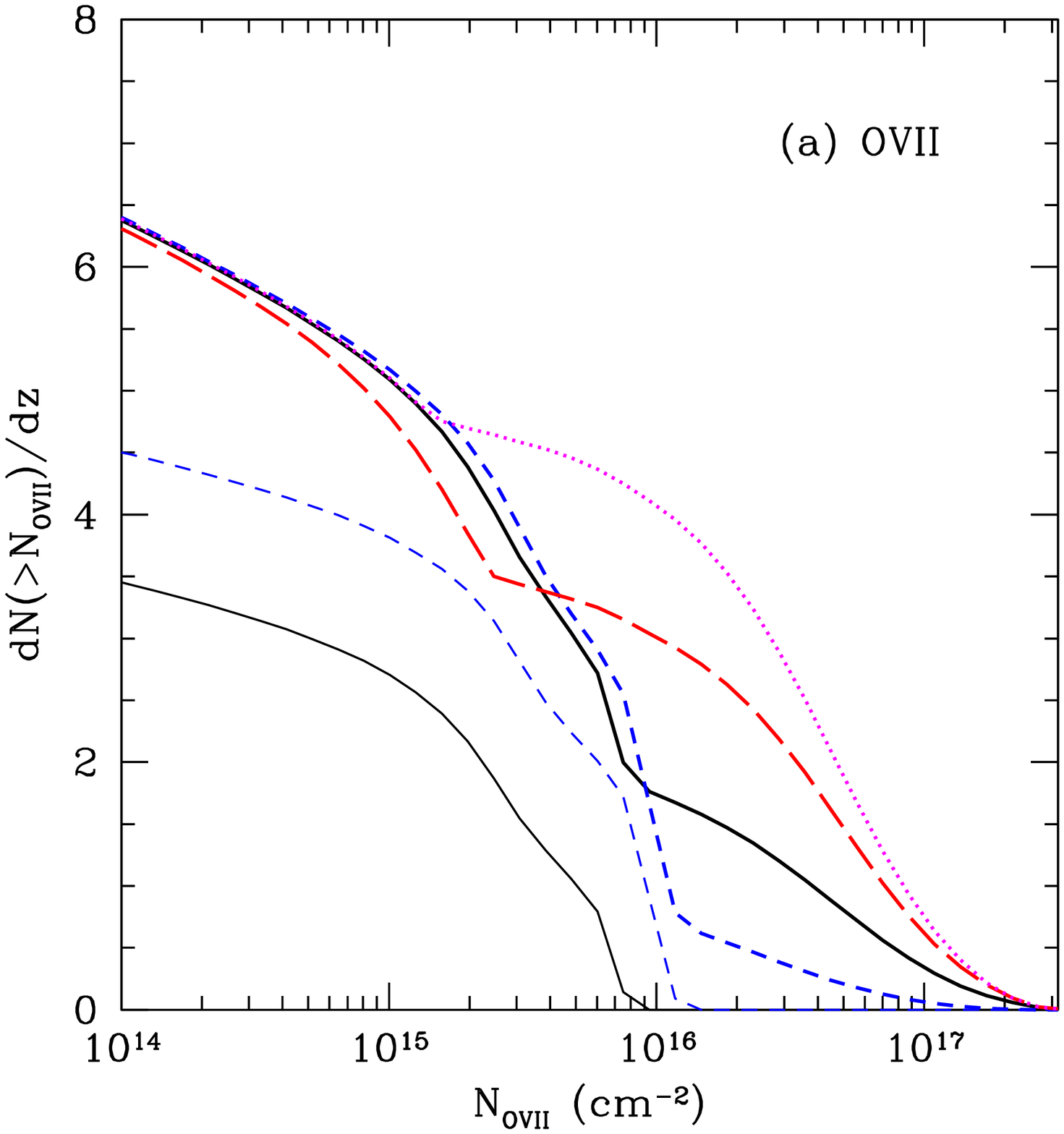}}%
\hspace{0.13cm}\resizebox{8cm}{!}{\includegraphics{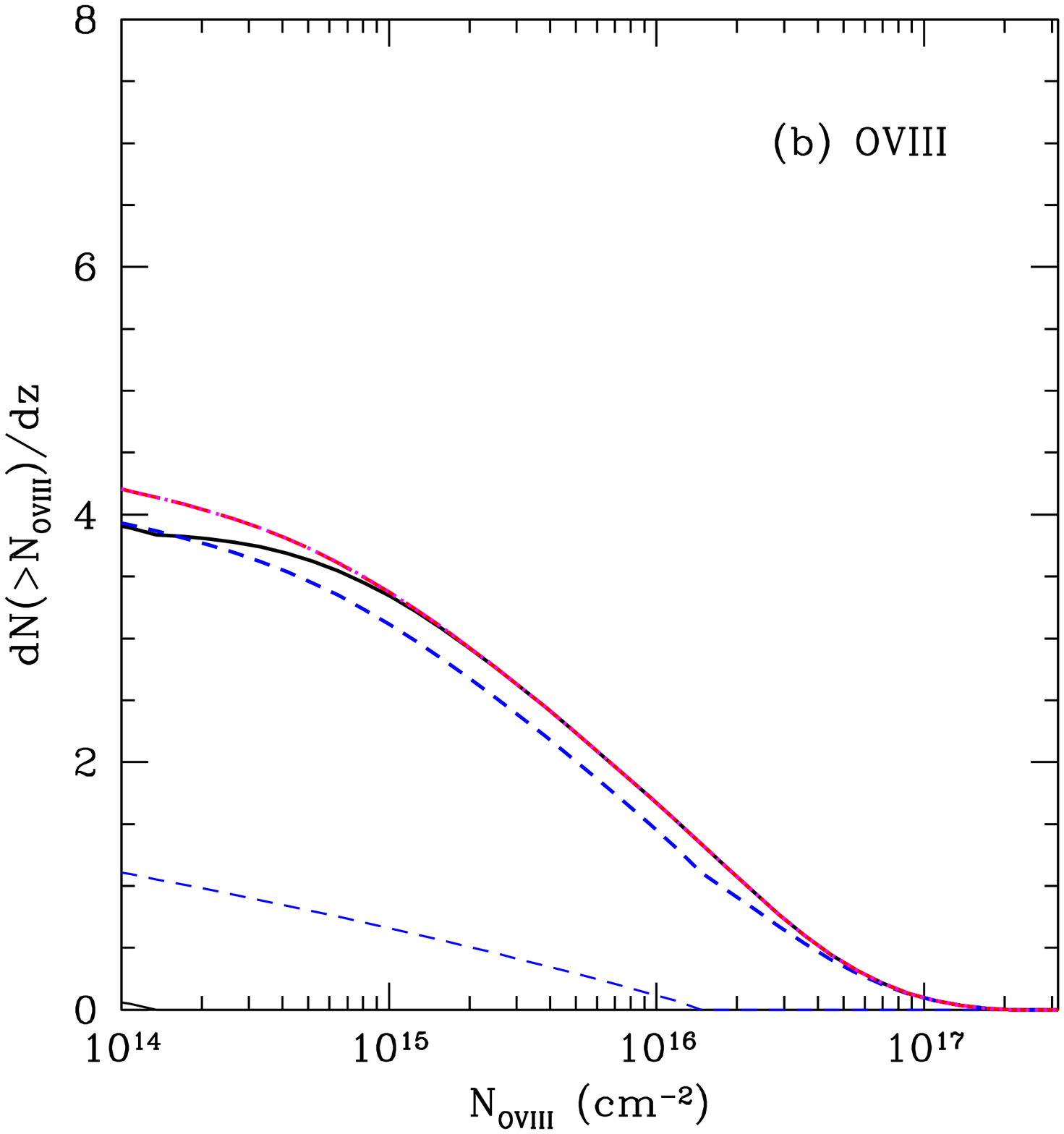}}%
\end{center}
\caption{Number of virialized absorption systems with column density larger than some threshold per unit redshift for the \oseven and \oeight ions (panels {\emph a} and {\emph b}, respectively) at $z=0$.  Long-dashed, solid, and short-dashed curves assume $Z=0.03,\, 0.1,$ and $0.3 \Zsun$ with $\Delta_{\rm cool}=200$.  The dotted curves assume $Z=0.1 \Zsun$ and $\Delta_{\rm cool}=50$. The thick curves show the total $\deriv N/\deriv z$, while the thin curves (plotted only for the solid and short-dashed cases) show the contribution from systems with $t_{\rm cool} < H_0^{-1}$. In panel {\emph b}, the dotted curve is coincident with the long-dashed curve and the thin solid curve is zero everywhere.}
\label{fig:vir_xray}
\end{figure*}
  
To compute the absorber statistics, we will follow the simple picture of \citet{white91} and divide the gas into two components.  First, if $t_{\rm cool}<H_0^{-1}$, we assume that it cools fully.  We evaluate the cooling time at a density $\Delta_{\rm cool}$ (usually 200) reflecting the typical density within a halo.  In this case, we can compute the column density of each system using the postshock cooling model
described in \S \ref{ncol}: each halo mass then corresponds to a single column density independent of the impact parameter (at least if we neglect orientation effects).  Equation (\ref{eq:cs}) gives the number of absorbers per unit redshift provided that we include only those halos hosting absorbers of
the specified column.  From Figure~\ref{fig:ionfrac}\emph{b}, it is obvious that at virial overdensities, cooling will be complete for halos with $T_{\rm vir} \la 10^6 \kel$.  This regime will thus be especially important for the \osix absorbers.

In the other regime, where $t_{\rm cool}>H_0^{-1}$, the gas settles into a quasi-equilibrium hot halo with $T \approx T_{\rm vir}$.   We could again use the model of \S \ref{ncol} to compute the corresponding columns.  However, the density structure is likely to be important for virialized halos, and reasonably well-motivated models for the profile do exist.  We will therefore use a simplified version of \citet{perna98}, in which we integrate along a line of sight through the halo to find the column density; obviously the column density does depends on the impact parameter in this case.  We will assume that a fraction $f_h=0.5$ of an object's baryonic mass is contained in a diffuse gaseous halo within $r_{\rm vir}$.  We will also assume that its density follows a beta-model distribution, $\rho \propto [1+(r/r_c)^2]^{-1}$, where $r_c$ is the core radius.  We will take $r_c = r_{\rm vir}/4 \equiv c_\beta r_{\rm vir}$, which is
approximately consistent with the profiles of luminous clusters \citep{jones99}.  We will assume isothermal gas, so the column density $\Nion$ along a line of sight with impact parameter $b$ is
\bqa
\Nion & = & f_{\rm ion} \, Z \, [{\rm O/H}]_\odot \, f_h \, \frac{2
  \Delta_c \, \bar{n}_H(z)}{3 \Omega_m \, g(c_\beta)} \,
\frac{r_c}{\sqrt{1+b^2/r_c^2}} \nonumber \\ 
& & \times {\rm tan}^{-1} \left(
\sqrt{\frac{r_{\rm vir}^2 + b^2}{r_c^2 + b^2}} \right),
\label{eq:vircol}
\eqa
where the metallicity is in solar units, $\Delta_c$ is the virialization overdensity in units of the critical density \citep{bryan98}, and $g(x) = x^2(1-x \, {\rm tan}^{-1} x^{-1})$.  In the slow-cooling limit, we have $\Nion \propto f_h Z$.  The number of systems per unit redshift is then the analog of equation~(\ref{eq:cs}), 
\bq
\frac{\deriv N(>\Nion)}{\deriv z} = \frac{\deriv r}{\deriv z} \int \deriv m \, n(m) \, \pi R^2(m,\Nion), 
\label{eq:dndzx}
\eq
where $R(m,\Nion)$ is the maximum impact parameter for a halo of mass $m$ within which the column exceeds $\Nion$.  

Thus, to compute the total number of absorbers from virialized halos, we begin by dividing halos into fast-cooling and slow-cooling regimes (this amounts to finding the halo mass above which $t_{\rm cool}$ exceeds the age of the universe).  For masses above this threshold, we approximate the halo as isothermal and use the halo mass profile of equation (\ref{eq:vircol}) to convert to column density as a function of impact parameter; the abundance is then given by equation (\ref{eq:dndzx}).  We need not specify a maximum halo mass, because $n(m)$ falls exponentially fast beyond the nonlinear mass scale.  For smaller masses, the column density is determined by the shock physics of equation (\ref{eq:no6heck}) \emph{independent} of impact parameter, and the corresponding cross-section is given by equation (\ref{eq:cs}) if we integrate only over those halos with sufficiently large columns.  In this case, the integral converges because small halos correspond to small virial temperatures for which highly-ionized oxygen is vanishingly rare (see Fig.~\ref{fig:nm}).

We show our results for  \oseven and \oeight in Figure~\ref{fig:vir_xray}.  The thick curves show the total
$\deriv N/\deriv z$, while the thin curves show the contribution from systems with $t_{\rm cool}<H_0^{-1}$.   Long-dashed, solid, and short-dashed curves assume $Z=0.03,\, 0.1,$ and $0.3 \Zsun$ with $\Delta_{\rm cool}=200$.  The dotted curves assume $Z=0.1 \Zsun$ and $\Delta_{\rm cool}=50$.
As expected from Figure~\ref{fig:cs}, the number of virialized absorbers is a few per unit redshift, well below the total frequency found in observations \citep{nicastro04} and   simulations \citep{fang02} but consistent with other analytic treatments of virialized objects \citep{perna98}.   Systems with $t_{\rm cool}<H_0^{-1}$ can make a significant contribution to the \oseven absorbers but are unimportant
for \oeight.  However, they are always relatively weak, with a characteristic $\Noseven \sim 10^{15.5} \colden$ (that is relatively independent of metallicity).  In contrast, hot halos can contain quite large columns, because the path lengths can be long.  The abundance of \oseven absorbers is larger than that of \oeight because halos near the nonlinear mass scale have $T_{\rm vir} \la 10^{6.5} \kel$, below the ionization threshold of \oeightns.  

Note that the number of high-column-density absorbers \emph{increases} if the metallicity decreases.  As the metallicity declines, the cooling time increases and less-massive halos retain their hot gas.  This makes up for the decreased column in any one absorber because $n(m)$ falls rapidly with mass.  This is in many respects an artifact of our simplified treatment:  for $t_{\rm cool} \sim H_0^{-1}$, we should include cooling to compute the detailed column densities.  

Here and below, we present results for $z=0$.  The statistics evolve only modestly to $z \sim 1$ (e.g., compare the two curves in Figure~\ref{fig:cs}).  Beyond that, the abundance falls rapidly because the nonlinear mass scale falls below $T_{\rm vir} \sim 10^6 \kel$.

\begin{figure}
\begin{center}
\resizebox{8cm}{!}{\includegraphics{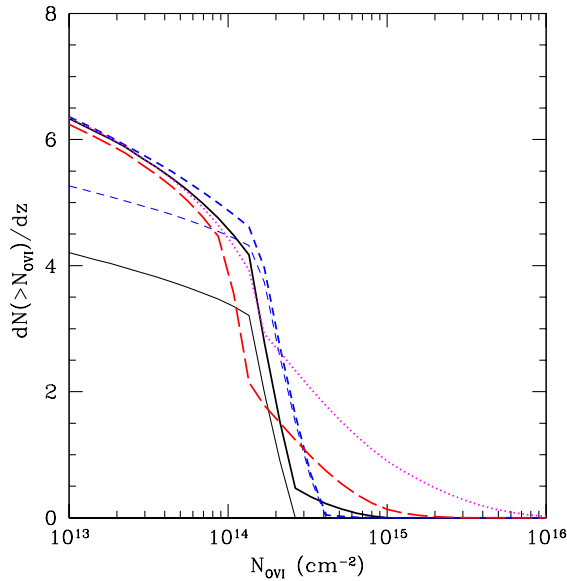}}\\%
\end{center}
\caption{Same as Fig.~\ref{fig:vir_xray}, except for \osixns.}
\label{fig:vir_o6}
\end{figure}

Figure~\ref{fig:vir_o6} shows analogous results for \osixns.  In
this case, the rapidly-cooling halos are even more important,
yielding a characteristic column density $\Nosix \sim 10^{14.5}
\colden$ (see Figure~\ref{fig:nm}).  The high-$\Nosix$ tail is
composed of halos just above the cooling threshold which still
have relatively large $\fosix$.  Thus it only appears if the
cooling time is long (i.e., $Z$ or $\Delta_{\rm cool}$ is small)
and is subject to uncertainties in our simplified cooling model.

As required by Figure~\ref{fig:cs}, virial shocks underpredict $\deriv \Nosix/\deriv z$ and $\deriv \Noseven/\deriv z$ by factors of several.  Clumping will further decrease the cross section.  If we place the shocked gas into $n_b$ blobs of radius $r<r_{\rm vir}$, only a fraction $f_{\rm cov} \approx n_b (r/r_{\rm vir})^2 < 1$ of lines of sight through each halo would host absorbers.  It could, however, increase the column densities.  In the slow-cooling case, we would have $\Nion \propto f_{\rm cov}^{-1}$, increasing the amplitude of the strong-absorber tail.  On the other hand, fast-cooling gas would be unaffected.

Finally, we note that we have allowed the gaseous halos in
groups to remain relatively dense and hot.  It is well-known
that such models overpredict the soft X-ray background
\citep{pen99,wu01}.  One solution is to introduce a feedback
mechanism that decreases the density of the intragroup medium
\citep{wu01,xue03}.  This would, of course, modify our
column-density distribution.  By slowing the cooling, the
principal effect would be to increase the amount of absorption.
It could also slightly increase the cross section if the gas is spread over larger distances.  However, simple geometry demands that such pre-heating scenarios will certainly not bring the abundance into agreement with observations; it is more likely to change the column densities by a modest amount.  In principle, the distribution of strong IGM absorbers could then be used as a probe of feedback mechanisms, though the many other uncertainties (such as metallicity and non-equilibrium effects) make this a difficult endeavor.

\subsection{Infalling Gas}
\label{infall}

To reproduce the large number of absorbers found in simulations and observations, we \emph{must} appeal to a population beyond virial shocks.  This is of course not surprising, because simulations contain large networks of IGM shocks.  To connect these shocks with collapsed objects, we will postulate that the temperature $T_{\rm sh}$ of IGM shocks surrounding each dark-matter halo depends (exclusively) on the halo mass.  One way to motivate such a picture is if the shocks occur during
infall onto existing structures.  \citet{furl04-sh} used this {\it ansatz}, together with spherical symmetry, to estimate the distribution of IGM shocks.  They argued that shocks could plausibly occur when an object breaks off from the cosmological expansion (at ``turnaround''), because that is the point at which the flows begin to converge.  Assuming strong shocks, the peculiar velocity at turnaround implies $T_{\rm sh} \approx T_{\rm vir}/4$.  The model roughly matches simulation estimates of the mass fraction and characteristic temperature of the WHIM.  Although clearly a simplification, this picture does describe many of the major features of the WHIM and provides a well-defined way to convert $n(m)$ to the distribution of $T_{\rm sh}$.  Furthermore, spherical infall occurs on quite large scales and could provide the cross section we need.  Turnaround occurs at $r_{\rm ta} \approx 3.3 r_{\rm vir}$ for self-similar spherical infall \citep{bert85} in an Einstein-de Sitter universe, so the shock networks could be quite large relative to collapsed objects.  Simulations also show networks of shocks surrounding bound structures at a few times the virial radii \citep{keshet03,nagai03}.  We will, however, leave the cross section of each infall shock free, because it must depend on the asymmetric, filamentary nature of the infall.  We will parameterize it as $\chi \pi r_{\rm vir}^2$.  For reference, the top axis of Figure~\ref{fig:nm} shows $T_{\rm sh}(m)$ in this infall model.

Because of the lack of data pertaining to IGM shocks, a detailed comparison of this model with observations is difficult.  However, \citet{rines03} have used galaxy redshifts to trace the infall regions of eight nearby clusters spanning about an order of magnitude in mass.  Although the galaxies obviously will not be subject to shocks, their structure should reflect the underlying gas distribution at least qualitatively.  Most importantly, \citet{rines03} found that $r_{\rm ta} \approx 3.5 r_{\rm vir}$ in all of their clusters.  The spherically-averaged density profiles fall as $\sim r^{-3}$--$r^{-4}$ at large radii, consistent with simulations \citep{hernquist90,navarro97}. This implies that the average density falls by about a factor of fifty between the virial radius and turnaround radius.  However, as we will show in \S \ref{sims} below, the density field is highly inhomogeneous within this region.  For simplicity, we will assume a constant density behind these shocks.  Finally, \citet{rines03} compute the velocity dispersion profiles.  Between $r_{\rm vir}$ and $r_{\rm ta}$, the dispersion typically declines by a factor of two or so.  This is nicely consistent with our assumption that $T_{\rm sh} \approx T_{\rm vir}/4$.

\begin{figure}
\begin{center}
\resizebox{8cm}{!}{\includegraphics{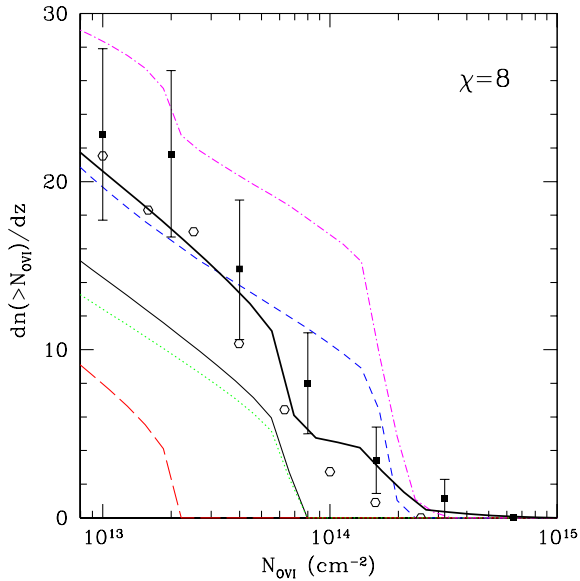}}\\%
\end{center}
\caption{Number of absorption systems from infall shocks for the \osix ion at $z=0$, assuming $\chi=8$. The long-dashed, thin solid, and short-dashed curves assume $Z=0.03,\,0.1$, and $0.3 \Zsun$, respectively, with $\Delta=10$.  The dot-dashed curve assumes $Z=0.1 \Zsun$ with $\Delta=100$.  The dotted curve is the same as the solid curve except that we take $T_{\rm sh}=T_{\rm vir}/6$.  The two sets of points show observed results (the solid squares are described in the text; the open hexagons are from \citealt{danforth05}).  The thick solid curve adds the contribution from virialized systems.}
\label{fig:o6igm}
\end{figure}

Given this prescription, we can compute the expected number of absorbers using the postshock cooling model of \S \ref{ncol}.  Here we will use this model for both fast-cooling and slow-cooling shocks, because (unlike for virialized halos) we have no well-motivated model for the density profile of IGM filaments.  However, we caution the reader that geometry (such as the thickness of filaments) could still ultimately determine the column densities.  
We also note that, for these modest densities $\Delta \sim 10$, the recombination times can approach or even exceed the age of the universe, so our assumption of local CIE in calculating the abundances may be problematic (see discussion in \S \ref{ncol}).  To compute $\deriv n/\deriv z$, we use equation (\ref{eq:cs}), with the integration range set by the column density corresponding to each halo mass.
Operationally, the calculation is similar to that described in \S \ref{virial}, except that equation (\ref{eq:no6heck}) applies to both fast and slow-cooling shocks.

\begin{figure*}
\begin{center}
\resizebox{7.5cm}{!}{\includegraphics{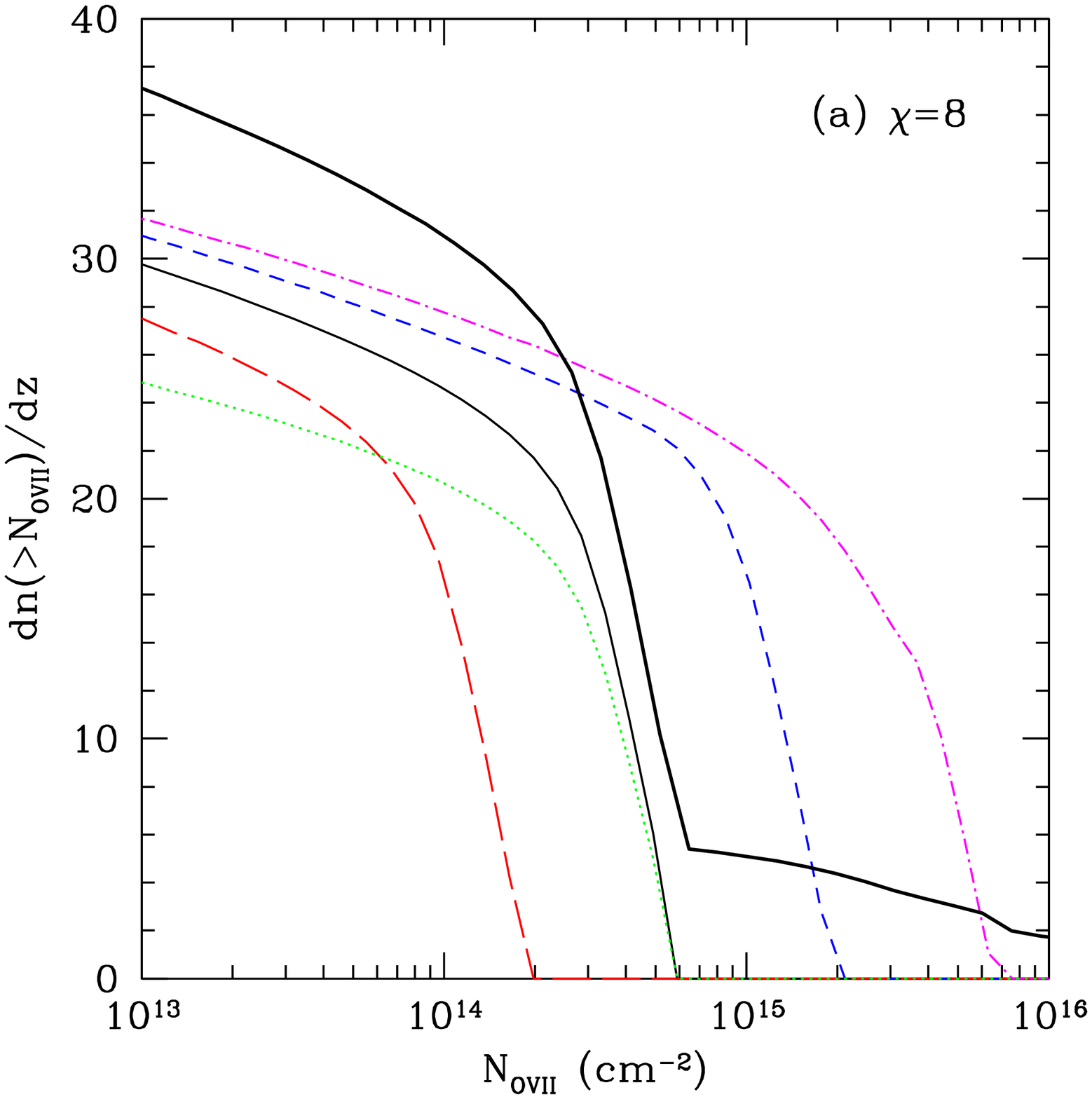}}%
\hspace{0.13cm}\resizebox{7.5cm}{!}{\includegraphics{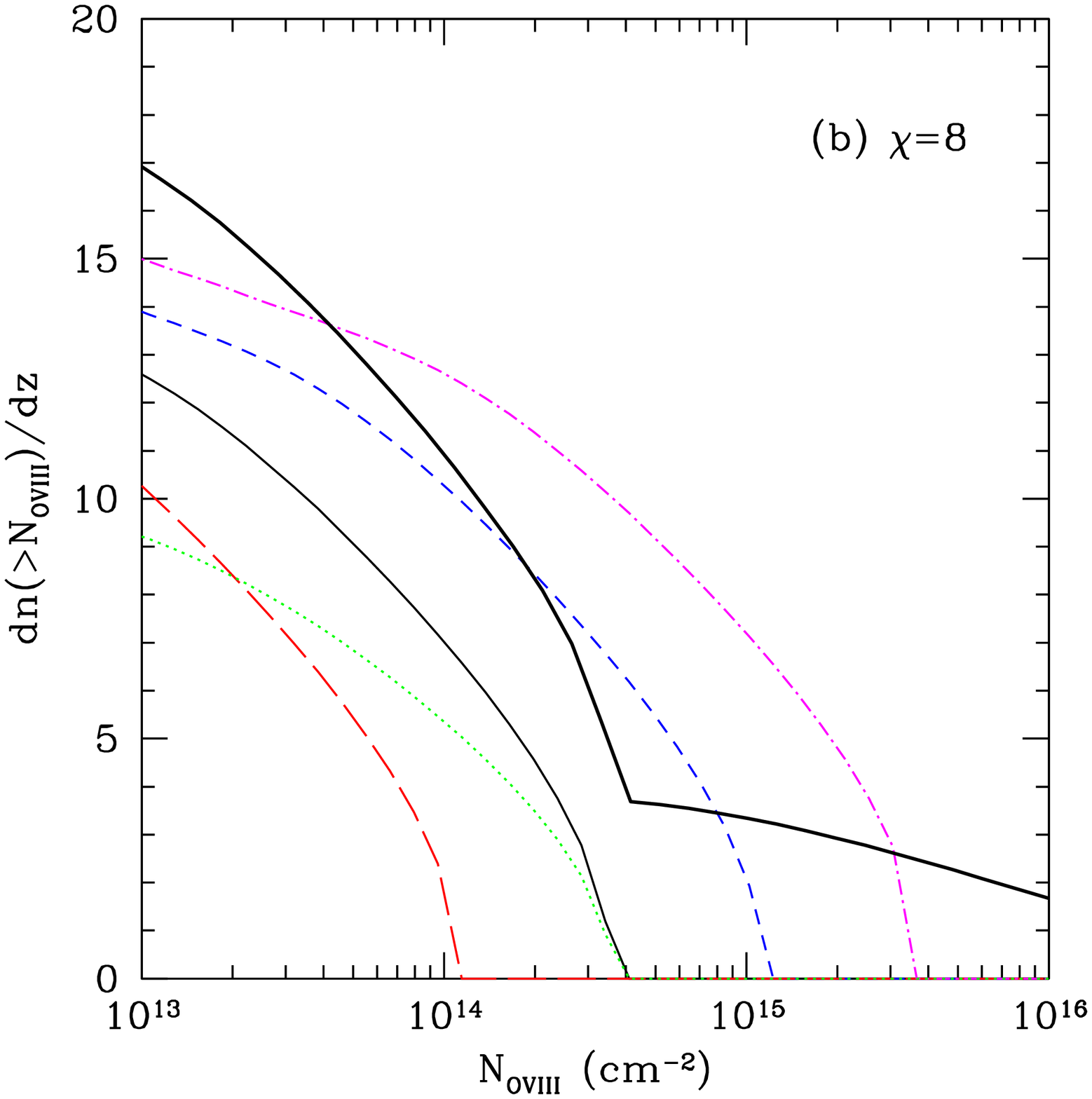}}%
\end{center}
\caption{Same as Fig.~\ref{fig:o6igm}, except for the \oseven and \oeight ions (panels {\emph a} and {\emph b}, respectively).  Note that the ordinates have different scales.}
\label{fig:xrayigm}
\end{figure*}

Given the simplifications of our model, a detailed comparison to observations is unwarranted.  However, we do wish to estimate the required $\chi$ as well as check whether the derived column densities are reasonable.  As such, we begin with \osixns, which is by far the best studied ion.  The filled squares in Figure~\ref{fig:o6igm} show a distribution compiled from several lines of sight in the literature \citep{tripp00,sembach01,richter04,prochaska04}; these lines of sight have the best-defined samples.\footnote{However, we have not attempted to account for the different detection limits of each survey.  As such, the statistics at $\Nosix \sim 10^{13} \colden$ may not be reliable.}  We have included all claimed detections (even those labeled ``tentative'' by each survey's authors).  The error bars assume Poisson statistics.  The open hexagons show the more systematic survey of \citet{danforth05}, who searched for \osix absorbers in 129 known Ly$\alpha$ absorbers along 31 lines of sight studied with the \emph{FUSE} satellite.  We refer the reader to that paper for a detailed discussion of the errors.  Note that the two datasets are reasonably consistent, though the \citet{danforth05} data do lie below our compilation by a small amount.

The curves in Figure~\ref{fig:o6igm} show the results from our model assuming $\chi=8$.
The long-dashed, solid, and short-dashed curves assume $Z=0.03,\,0.1$, and $0.3 \Zsun$, respectively, with $\Delta=10$.  The dot-dashed curve assumes $Z=0.1 \Zsun$ with $\Delta=100$.  The dotted curve is the same as the solid curve except that we take somewhat weaker shocks, with $T_{\rm sh}=T_{\rm vir}/6$.  The thick solid curve adds the contribution from virialized systems for a better comparison with the observations.  

In all cases, the number of \osix absorbers falls sharply at some characteristic column density.  This is simply the maximum allowed for fast-cooling shocks.  In contrast to virial shocks, however, the densities are typically small enough that the maximum column density does depend on the metallicity and physical density (unless $Z \sim 0.3 \Zsun$ or $\Delta \sim 100$, in which case the cooling times begin to fall below the age of the universe).  Below the characteristic maximum, the number of absorbers rises rapidly because $\fosix$ is so sensitive to the shock temperature.  We thus expect infall shocks to contribute only to the range $\Nosix \la 10^{14} \colden$.  Note that the sharp rise in the dot-dashed curve at $\Nosix \approx 2 \times 10^{13} \colden$ occurs because of the plateau in $\fosix$ at $T \sim 10^6 \kel$ (see Fig.~\ref{fig:ionfrac}).  Of course, the maximum $\Nosix$ does not decrease for $T_{\rm sh}=T_{\rm vir}/6$ because that has no effect on the cooling.  It decreases $\deriv N/\deriv z$ by a small amount because it shifts the halo mass corresponding to each $T_{\rm sh}$.

Thus, if all the known absorbers are to be attributed to shocks (either infall or virial), we find that $\chi \sim 8$; i.e., the infall shocks subtend an area eight times larger than the halo itself.  While large, this is smaller than the expectation from pure spherical infall so may not be unreasonable.  
Moreover, as we will discuss in \S \ref{photo} below, a number of the observed systems appear to be photoionized and have a different origin, which would allow the covering factor of infalling gas to decrease by about a factor of two.  This would decrease the thin curves in Figure~\ref{fig:o6igm} by the same factor:  note that we would still expect a sharp decline near the characteristic maximum column density of the shocks.  Because stronger systems only arise in virial shocks, the covering fraction within virialized objects can be separately calibrated from the abundance of strong absorbers (though of course that could also be a function of halo mass).  In the simplest model, a significant population of photoionized absorbers would fix $\chi \approx 4$, allowing IGM absorbers to be more filamentary, while leaving the virial shock spherical.  In any case, we find it reassuring that a reasonable choice for $\chi$ can simultaneously match the observed abundance and column-density distribution, at least roughly.

We repeat the same exercise for \oseven and \oeight in Figure~\ref{fig:xrayigm}.  We have again fixed $\chi=8$.  First, note that the number and characteristic columns of \oseven absorbers are much larger than for \osix absorbers.  This follows easily from Figure~\ref{fig:nm}, because \oseven dominates over a much larger temperature range (including the \osix peak).  On the other hand, the maximum $\Noseven$ is poorly constrained: because the cooling times always exceed the age of the universe, the cutoff is proportional to metallicity and density.  Interestingly, \oseven drops off steeply at this maximum and is nearly flat below it, so most absorbers of a given metallicity and density should cluster around one particular column.  This occurs because every system cools over $H_0^{-1}$ and $\foseven$ is nearly independent of temperature.  In reality, the cutoff will be smeared by the range of metallicity and density as well as geometry.  For X-ray lines, the comparison to observations is much more difficult given current technology.  There is only one well-studied line of sight (with secure detections of two absorbers); formally this implies $\deriv N/\deriv z(\Noseven \ga 7 \times 10^{14} \colden) = 67^{+88}_{-43}$ \citep{nicastro04}.  Our model is well within this range.  More interesting is to note that the observed columns ($7 \times 10^{14}$ and $10^{15} \colden$) are both near the cutoff and appear to require $\Delta \ga 30$ or $Z \sim 0.3 \Zsun$, still well within reasonable bounds.  Simulations suggest $\deriv N/\deriv z(\Noseven \ga 7 \times 10^{14} \colden) \sim 10$--$30$, close to our model with $\chi=8$ \citep{fang02,chen03}.  They do not show as severe a break as our model, probably because of scatter in the metallicity and geometric effects.

Figure~\ref{fig:xrayigm}{\emph b} shows that the expected number of \oeight absorbers is about half that of \osevenns, although the column densities can be comparable.  \oeight is rarer simply because the characteristic shock temperature is $\la 10^{6.5} \kel$ \citep{furl04-sh}, so there is not as much gas with large $\foeight$.  As a result, virialized systems make a larger relative contribution.  Also, note that \oeight does not show a sharp cutoff; this is because $\foeight$ varies rapidly with temperature.

\section{Comparison to Simulations}
\label{sims}

We have shown that the spherical-infall model described in the previous section can explain many of the key properties of the absorbers.  However, it also makes a number of important assumptions and  simplifications.  Here we compare our model to simulations in order to highlight its strengths and weaknesses.

\subsection{The Cosmological Simulation}
\label{sim-summ}

We use the $z=0$ dark-matter, stellar-particle, gas-temperature and density outputs of a large scale $L=25h^{-1}$Mpc cosmological hydrodynamical  simulation, with $768^{3}$ fluid elements and $384^{3}$ dark-matter particles (each with mass  $2 \times 10^{7}h^{-1}M_{\odot}$). The code structure is similar to that in \citet{cen92a, cen92b} with some significant changes \citep{cen00}. It is described in detail in \citet{nagamine01}.  The chosen cosmology is the ``concordance model" \citep{wang00}, a flat low-density ($\Omega_{m} = 0.3$) universe with a cosmological constant ($\Omega_{\Lambda} = 0.7$), and $h=0.67$.  The baryon density (originally $\Omega_b=0.049$) has been scaled to $\Omega_{b} = 0.046$ to match the value used in our analytical model. The galaxy particles are formed through the recipe described in the Appendix of \citet{nagamine00}. Each stellar particle is assigned a position, velocity, formation time, mass, and metallicity at birth. The stellar particle is placed at the center of the cell with a velocity equal to the mean velocity of the gas in the cell. It is then followed by a particle-mesh code as a collisionless particle interacting gravitationally with the dark matter and the gas. We use the galaxy and group position information obtained from these stellar particles  by \citet{nagamine01}.

We will primarily be concerned with the IGM around bound objects. We use the dark-matter halo parameters for this simulation obtained by \citet{nagamine01} with the HOP grouping algorithm of
\citet{eisenstein98}. We then search for the halos that are most closely coincident ($\Delta L \leq 68\, h^{-1}\,\kpc$) with the galaxy-particle groups and use the properties of these dark-matter halos. The virial radius is obtained from halo mass using equation (24) of \citet{barkana01}.  Around the center of each halo, we obtain the volume-averaged radial distributions of matter, temperature, and metals in the inner regions abutting the galaxy-particle locations in the $z = 0$ output:
\begin{equation}
\frac{\int d^{3}x \rho(x) Q(x)}{\int d^{3}x \rho(x)},
\end{equation}
where $Q(x)$ is the parameter value ($\Delta$, $T$, $f_{\rm ion}$, etc.) in the volume element $d^{3}x$.
For the sake of transparency, we do not directly use the metallicities reported by the simulation.  Instead we consider two metallicity prescriptions. The first scales metallicity with density as described by \citet{croft01}, $Z = 0.005 \Zsun \times \sqrt{n/\bar{n}}$, with a maximum metallicity of $Z =
0.33 \Zsun$. The second prescription simply uses a constant metallicity $Z = 0.1\,Z_{\odot}$. In order to study the \osix, \oseven, and \oeight abundance distributions, we compute the ionization fraction as a
function of  temperature as illustrated in Figure~\ref{fig:ionfrac}\emph{a}.  With $f_{\rm ion}$ for each cell, plus the density and metallicity, we compute the total ionic abundance relative to hydrogen.  We scale the distributions by the virial radius corresponding to each halo and follow them out to five virial radii. We obtain the volume-averaged distributions for several virial-radius bins, $r_{\rm vir}=240$--$300$, $360$--$480$, and $600$--$720 \kpc$; these correspond to roughly (1--2), (3--7), and (14--24) $\times 10^{12} M_\odot$.

\subsection{Absorbers Around Virialized Halos}
\label{sim-vir}

\begin{figure}
\begin{center}
\resizebox{8cm}{!}{\includegraphics{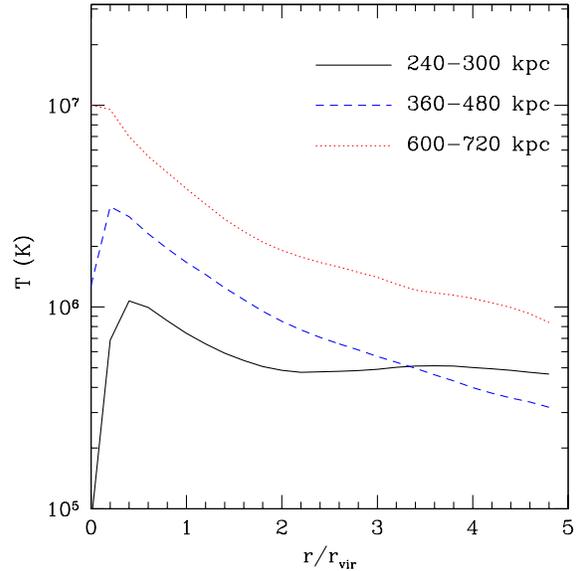}}\\%
\end{center}
\caption{Spherically-averaged temperature profiles around virialized halos in a cosmological simulation, where all distances are scaled to the virial radius.  The three curves show different mass ranges.}
\label{fig:tsim}
\end{figure}

The key {\it ansatz} in our infall-shock model is that the characteristic IGM temperature surrounding a collapsed object increases with its mass, even on scales much larger than $r_{\rm vir}$.  Figure~\ref{fig:tsim} shows spherically-averaged temperature profiles around halos in the simulation (normalized by $r_{\rm vir}$).\footnote{At $r \ll r_{\rm vir}$, the temperature is strongly affected by star formation and feedback, so we will ignore that portion.}   We see that isolated galaxies tend to lie in cooler regions of the IGM than massive groups, even at several times $r_{\rm vir}$.    The trend appears to break down at $r \ga 3 r_{\rm vir}$ for the smallest objects.  This is in a temperature range where cooling can be significant, which may affect the results.  Also note that there is no characteristic radius at which the temperature increases, as would be expected for pure spherical infall.  Thus the shocks appear to be complex and distributed, with a range of temperatures.  The typical temperature is, however,  a few times smaller than $T_{\rm vir}$, consistent with our model.  

\begin{figure}
\begin{center}
\resizebox{8cm}{!}{\includegraphics{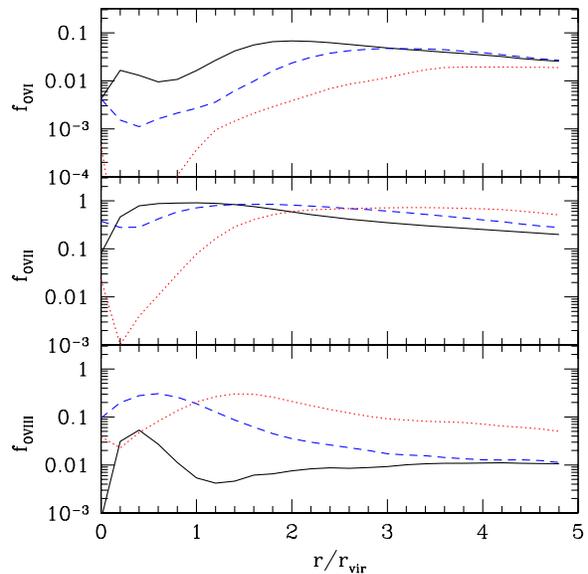}}\\%
\end{center}
\caption{Spherically-averaged ionization fractions around virialized halos in a cosmological simulation, where all distances are scaled to the virial radius.  The top, middle,and bottom panels are for the \osixns, \osevenns, and \oeight ions.  In each panel, the three curves show different mass ranges, as in Fig.~\ref{fig:tsim}.}
\label{fig:fionsim}
\end{figure}

Figure~\ref{fig:fionsim} shows that it may be more accurate to
associate different oxygen ions with different halo masses.  The
three panels show the
spherically-averaged ionization fractions for each of our ions
as a function of normalized radius.  The curves correspond to
the same mass ranges shown in Figure~\ref{fig:tsim}.  We see
that \osix tends to be associated with the smallest-mass
objects, although even these halos are too hot to sustain a
large $\fosix$ inside $r_{\rm vir}$.  On the other hand, \oeight
is most prevalent around the highest-mass objects (although not
within the virialized regions, again because the gas is too hot)
and is extremely rare around the low-mass objects.  Finally,
\oseven appears in all systems because it dominates over such a
wide temperature regime; only the virialized gas in extremely
massive objects is hot enough to destroy \oseven.

\begin{figure}
\begin{center}
\resizebox{8cm}{!}{\includegraphics{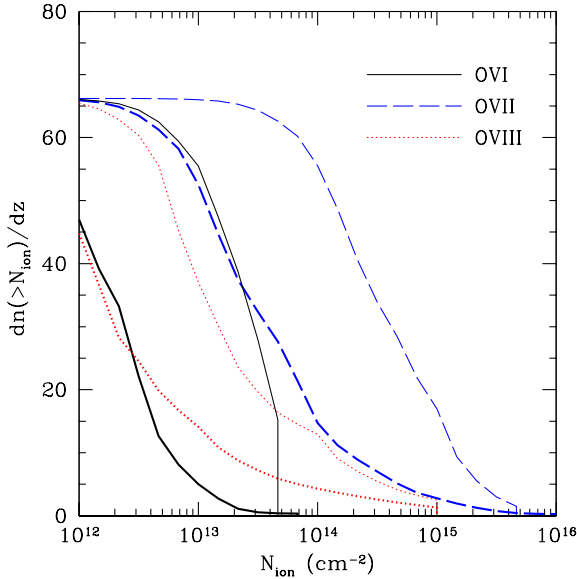}}\\%
\end{center}
\caption{Absorber statistics for spherically-averaged profiles taken from the simulations.  The different line styles show the three different ions.  The thin curves assume a uniform $Z=0.1 \Zsun$ metallicity; the thick curves use a density-dependent metallicity prescription (see text).}
\label{fig:dndzsim}
\end{figure}

Thus, we can be confident in expecting highly-ionized absorbers to lie near massive collapsed objects.  However, the simulations also show that a simplified spherical model is not accurate.  In Figure~\ref{fig:dndzsim} we have computed $\deriv n/\deriv z$ for each ionic absorber using our spherical profiles.  We have used equation (\ref{eq:dndzx}) with the column densities determined by integrating through the average profiles.  In each case we truncate the profile at $5 r_{\rm vir}$.  The dotted, dashed, and solid curves show the expected statistics for \osixns, \osevenns, and \oeightns, respectively.  First consider the thin curves, which assume a constant $Z=0.1 \Zsun$ metallicity (as in our fiducial analytic model).  We see $\deriv n/\deriv z \sim 65$ for each ion at sufficiently small column densities.  This is an artifact of truncating the integrating at $5 r_{\rm vir}$.  The profiles in Figure~\ref{fig:fionsim} tend to approach nearly constant values at large radii, so for large impact parameters the column density is essentially proportional to the maximum radius; only well above this turnover can the column densities be trusted.

In that range, \oseven and \oeight do not present dramatically
different pictures from the analytic model.  Because $\foseven$
is nearly temperature-independent over a broad range, we might
expect spherical averages to do a passable job with that ion.
Indeed, the number density does agree roughly with our model and
with more detailed simulation analyses \citep{fang02,chen03}.
But it is certainly not so good that we can claim support for
our model.  \oeight is considerably worse, and the \osix results
do not appear to have converged and disagree strongly with our
model.  This is also not surprising, because \osix is often in a
fast-cooling regime and is so temperature sensitive that
averaging over large volumes artificially suppresses its
importance.  Ray tracing appears vital to predict the \osix
absorber abundance, as would be expected in a shock-cooling
model.

The thick curves in Figure~\ref{fig:dndzsim} assume the density-dependent metallicity distribution described above.  The number of absorbers declines sharply in this scenario:  at large radii the mean physical densities are near the cosmic mean, so the overall metallicity is small.  As a result, high column densities require small impact parameters.  One fortunate consequence is that the curves converge better.  However, much more significantly, the number of absorbers is far below the result of more complete simulation analyses \citep{cen01,fang01,fang02,chen03}.  Thus, the spherical averages do not accurately describe the absorbers in simulations.  Of course, the reason is the clumpiness of the IGM.    

\begin{figure*}
\begin{center}
\resizebox{14cm}{!}{\includegraphics{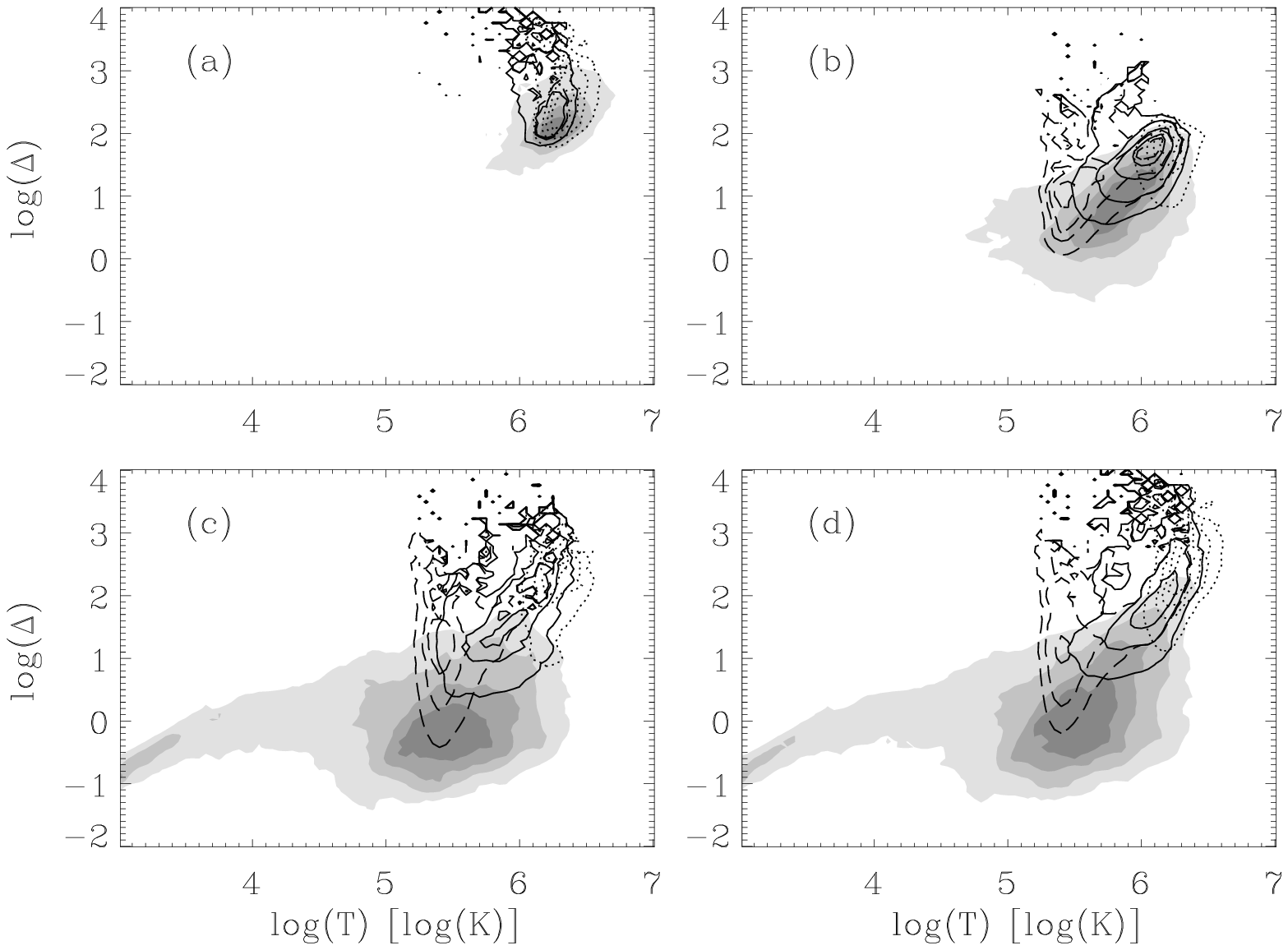}}\\%
\end{center}
\caption{Contour plots, in the density-temperature plane, of the volume-weighted (grey-scale), $\nosix$-weighted (dashed), $\noseven$-weighted (solid), and $\noeight$-weighted (dotted) distribution of gas within 0--$1 r_{\rm vir}$ (panel {\emph a}), $1$--$3 r_{\rm vir}$ (panel {\emph b}),  $3$--$5 r_{\rm vir}$ (panel {\emph c}), and 0--$5 r_{\rm vir}$ (panel {\emph d}) of halos with $r_{\rm vir}=360$--$480 \kpc$ (or $M=3$--$7 \times 10^{12} M_\odot$).  The contours enclose $0.95$, $0.75$, $0.5$ and $0.25$ of the total volume or ion number.}
\label{fig:rhoTsim}
\end{figure*}

Figure~\ref{fig:rhoTsim} shows phase diagrams of the IGM gas surrounding halos with $M=3$--$7 \times 10^{12} M_\odot$ in four radial slices. The $\nosix$, $\noseven$, and $\noeight$-averaged contours were obtained using the \citet{croft01} prescription for the metallicity. Hotter gas tends to sit at higher densities (and hence occupy a small volume).  This correlation holds even at $\Delta \sim 1$--$10$, well below virial overdensities. Moreover, the mean temperature and density decrease and the scatter in the relation increases as one moves away from the central galaxy.  However, regardless of radius there is a non-negligible fraction of gas at high temperatures comparable to $T_{\rm vir}$.  This gas (which also has a relatively large density) is presumably responsible for most of the oxygen absorption.  Thus, as can be seen from Figure~\ref{fig:rhoTsim}{\emph d}, the higher-ionization states will tend to probe the densest filaments (and the most enriched gas, on average) around each structure, in addition to the largest structures.  Meanwhile most of the volume is filled with cooler, low-density gas that has a small metallicity given our prescription. The averaging procedure smooths the high-density gas, boosting the column density along all lines of sight under a constant metallicity.  But when the metallicity is density-dependent, the tenuous gas is so metal-poor that the dense clumps essentially disappear during averaging. These clumps are reflected in the fragmentary nature of the density-averaged contours at higher $\Delta$. As one moves outward, they represent an ever decreasing fraction of the volume but continue to contribute significantly to $\noseven$, $\noeight$, and even $\nosix$. The distribution is highly complex:  there is a wide range of temperatures for any given density (spanning about a decade for $\Delta \sim 10$), which reflects the non-uniformity of the infall process. 

How can we reconcile the apparent success of our analytic
treatment with the clear failure of spherical averaging in
simulations?  The key reason is that, with the exception of hot
virialized halos, we did not use spherical geometry in the
calculation; instead we used only the infall velocities and a
cooling time (capped at $H_0^{-1}$).  The physics of shock
cooling then predicted the column densities without recourse to
the gas distribution.  Thus we only need modify our picture to
describe a network of inhomogeneous shocks with total cross
section $\pi \chi r_{\rm vir}^2$ rather than a single coherent
structure.  Of course, it is still an open question whether our
model provides an accurate description of these networks even on a
mean level (specifically the shock temperature).  In the future,
quantitative treatments must also consider the importance of the scatter seen in
Figure~\ref{fig:rhoTsim}.

\section{Photoionized Gas}
\label{photo}

To this point, we have focused on systems that are cooling after
being collisionally ionized by a
shock.  In simulations, such systems make up only about half of
the \osix absorbers \citep{cen01,fang01}.  The others are
photoionized; such systems are rare for the strong absorbers but
dominate at  $\Nosix \sim 10^{13} \colden$.  In the simulations,
these absorbers are typically cooler and lower density than
collisionally-ionized systems.  Modeling of the observed \osix
systems also suggests that about half are photoionized (e.g., \citealt{prochaska04}).
However, this interpretation is subject to uncertainties in the
relative metal abundances, the ionizing background, and whether
the medium is multiphase.  Reliably separating this population
is crucial because it typically has $T \sim 10^4 \kel$, which
means that these systems are not part of the canonical WHIM.  We
have shown in \S \ref{ncol} (see Fig.~\ref{fig:nm}) that
postshock cooling predicts that individual \osix absorbers
should have \oseven columns about an order of magnitude larger
than $\Nosix$.  This is larger than $\fosix/\foseven$ in CIE at
the \osix peak because most shocks begin at higher temperatures
and cool into the \osix range, so \oseven probes a longer path
length through the postshock region.  We now calculate the
associated \oseven columns if photoionization dominates.

We have again used Cloudy (version 94, \citealt{cloudy}) to compute $\fosix$ and $\foseven$ assuming  ionization equilibrium with an input radiation background.  We constructed a grid in density and temperature with spacings $\Delta \log n_H = 0.5$ for $-7 < \log n_H < -3$ (in units of cm$^{-3}$) and $\Delta \log T=0.1$ for $3 < \log T < 7$ (in K).  We use the $z=0$ \citet{haardt01} ionizing
background normalized to a total ionizing rate $\Gamma_{-12} = 0.084$ (in units of $10^{-12} \secinv$).  This roughly matches the observational estimate of \citet{dave01}.

\begin{figure}
\begin{center}
\resizebox{8cm}{!}{\includegraphics{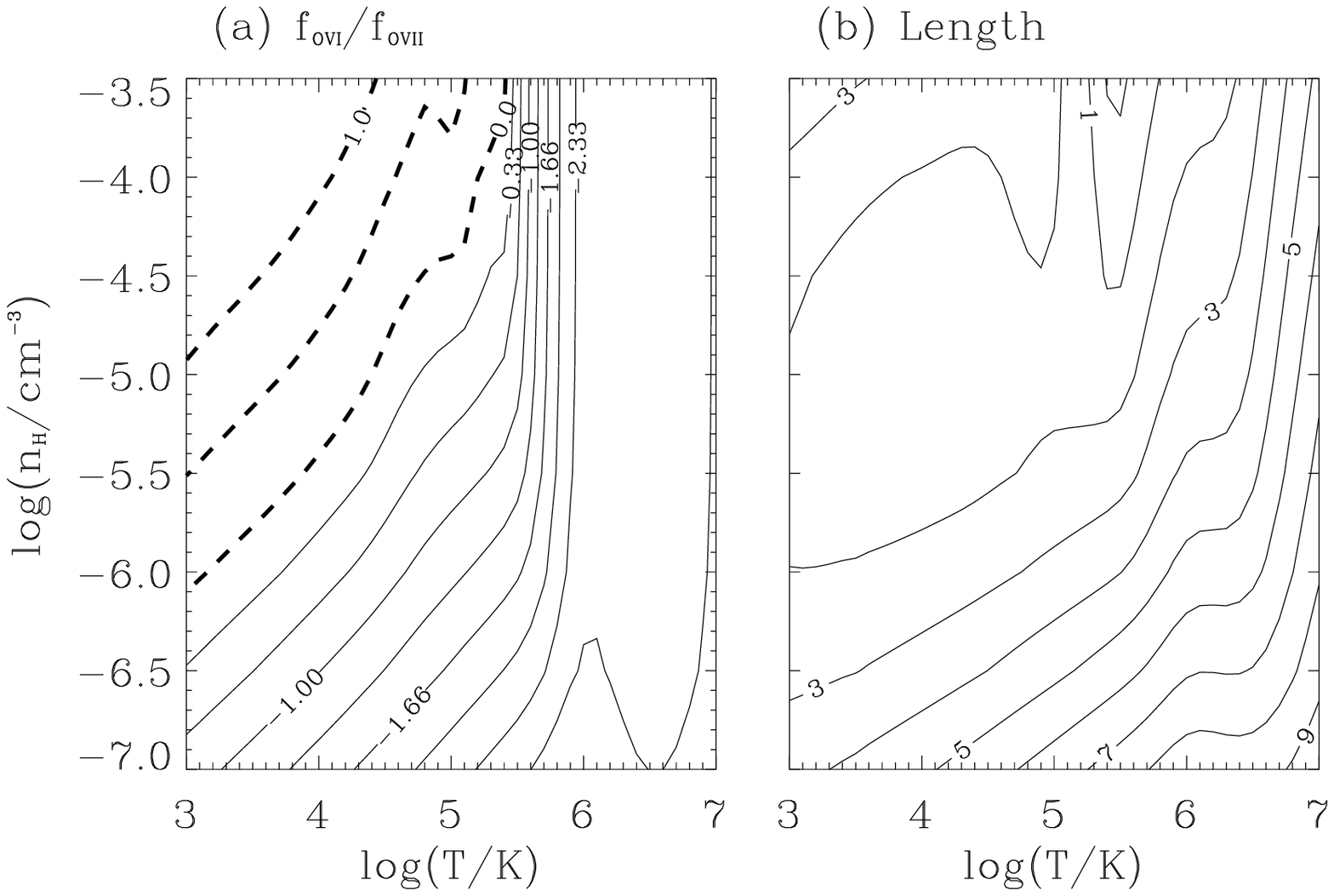}}\\%
\end{center}
\caption{\emph{(a)}: Contours show $\fosix/\foseven$ as a function of gas density and temperature, assuming photoionization equilibrium in a $z=0$ \citet{haardt01} ionizing background with $\Gamma_{-12}=0.084$. The dashed contours indicate $\fosix/\foseven>1$ while the solid indicate $\fosix/\foseven<1$; they are labeled with the logarithm of the ratio.  \emph{(b)}: The path length required for a total column density $\Nosix = 10^{13} \colden$ if the absorber has $Z=0.1 \Zsun$. The contours are labeled with the logarithm of the distance in kpc.}
\label{fig:hm01}
\end{figure}

Figure~\ref{fig:hm01} summarizes our results.  Panel \emph{(a)} shows the logarithm of the ratio $\fosix/\foseven$.  Dashed (solid) contours indicate $\fosix/\foseven>1$ $(<1)$.  To delineate those regions of
phase space where we could expect \osix absorbers in the IGM, panel \emph{(b)} shows the logarithm of the path length (in kpc) that would be required to produce an absorber with $\Nosix = 10^{13} \colden$ if $Z=0.1 \Zsun$.  Systems with $L \la 100 \kpc$ are confined to gas with $T \la 10^{5.5} \kel$ and $n_H \sim 10^{-6}$--$10^{-4} \cmden$ or to relatively dense gas lying close to the ionization temperature of
\osixns.  In the former case, we find that $\fosix/\foseven \ga 1$, unless the gas has $T \sim 10^{5.45} \kel$ (in which case collisional processes dominate and it is part of the WHIM) or is extremely tenuous and spatially large.  Thus, photoionized absorbers have smaller associated X-ray absorption than in the postshock cooling model and comparison of  $\fosix/\foseven$ offers a powerful test of the nature of the absorbing gas.  Upper limits of $\Noseven \sim 3 \Nosix$ can robustly identify those absorbers with $T < 10^5 \kel$ and separate WHIM systems from cooler photoionized gas.

This conclusion holds even if the ionizing background has a
different amplitude or shape.  The ratio $\fosix/\foseven$ is
nearly independent of $\Gamma_{-12}$.  The required path length
for low-density cool gas increases slightly as $\Gamma_{-12}$
increases, making tenuous absorbers more unlikely.  The shape of
the metagalactic background matters little since the ionization potentials of O{\sc V} and \osix are so closely spaced (114.2 eV and 138.1 eV, respectively).  We would have to introduce a (physically unmotivated) sharp feature to the background in this energy range to substantially change the ion ratios.  We have verified this with the \citet{haardt96} ionizing background, which includes only quasars and is significantly harder than the \citet{haardt01} background (which also includes galaxies).  The corresponding Figure~\ref{fig:hm01} looks identical.

It is worth emphasizing that $\fosix/\foseven$ is less subject
to modeling uncertainties than constraints that
use only UV lines (e.g., \citealt{prochaska04}) because it depends only on a single atomic species.  We also note that other ionization states could be useful.  For example, shocks that produce \oeight have $t_{\rm cool} \gg H_0^{-1}$ and should not contain substantial \osix.  Thus a large measured $\fosix/\foeight$ would help to identify photoionized systems.  Note that \citet{heckman02} imply that large columns of \osix and \oeight can coexist in the same absorber.  This is because they implicitly assume that the material cools fully; while reasonable for Galactic systems, this limit is not relevant for IGM absorbers.

Finally, \citet{heckman02} also pointed out that the linewidth---column-density relation offers another way to discriminate between photoionization and postshock cooling.

\section{Discussion}
\label{disc}

In this paper, we have attempted to elucidate the physical origin and properties of IGM oxygen absorbers (\osixns, \osevenns, and \oeightns) with a set of simple analytic arguments.  While simulations appear to predict approximately the correct number of absorbers \citep{cen01,fang01,fang02,chen03}, such studies have only examined the physical origin of the absorbing gas in a cursory fashion.  We first argued that the typical maximum column density of \osix absorbers ($\Nosix \sim 10^{14} \colden$) follows naturally from postshock cooling \citep{heckman02}, suggesting that at least some absorbers are generated by structure-formation shocks.  We then argued that, if these shocks are associated with bound structures, we must have a (more numerous) population of significantly hotter shocks around larger objects.  These shocks can be traced by \oseven and \oeightns; the former is more useful because it dominates over such a broad temperature range.  Moreover, shocked \osix systems must all have strong \oseven absorption.  This contrasts with photoionized \osix absorbers, which should have \oseven columns not much larger than the \osix columns.  

We showed that a simple picture with virial shocks and ``infall'' shocks can account for the observed number density of absorbers if each bound halo has a total cross section several times larger than $\pi r_{\rm vir}^2$.  
The inferred cross section depends on the fraction of observed \osix systems that are photoionized; if it is negligible, we find $\chi \approx 8$.  If about half of the weak absorbers are photoionized (as suggested by some recent data; \citealt{prochaska04}), $\chi \approx 4$; however, this does not necessarily affect the predicted abundance of strong absorbers, which mostly arise from virial shocks.
In our picture, the temperature of the infall shock increases with the central object's mass \citep{furl04-sh}.  Comparison to numerical simulations shows that these shocks are complex, with dense clumps of metal-enriched shocked gas embedded in a tenuous weakly-absorbing medium.  They do confirm our key assumption, the association of hotter shocks with more massive systems (albeit with substantial scatter).  In the future, we hope that more detailed comparisons to simulations will test other elements of our picture.

One of our most interesting results is that, while \oseven and \osix can be associated with the same shocked system (even neglecting multiphase media from thermal instabilities), they do not necessarily sample the same physical volume.  As the gas flows away from the shock, it cools over a range of temperatures.  \oseven is presumably associated with gas closer to the shock, while \osix probes more distant gas.  This calls into question modeling of the absorbers with single-temperature CIE gas.  Figure~\ref{fig:nm} shows that gas with a more or less constant $\Noseven$ can correspond to a wide range of $\Nosix$ because of the range of temperatures through which the gas element cools, even if it remains near equilibrium at any single temperature (which may not be a valid assumption; \citealt{dopita96}).  The crucial point is that gas with $\Delta \sim 10$--$100$ and $T \sim 10^5$--$10^7 \kel$ has a cooling time comparable to (though often a few times larger than) the Hubble time, so the temperature varies across the system.

The association of oxygen absorbers with virialization and
infall shocks obviously implies that these systems must be
spatially correlated with galaxy groups and clusters.  This is
not unique to our picture; any model that associates absorbers
with large-scale shocks in the cosmic web must show a similar
correlation.  There appears to be a relatively strong
correlation between \osix absorbers and galaxy groups:  most
lines of sight show galaxy overdensities coincident with the
absorber locations \citep{tripp00b,savage02,richter04}, although
it has not been statistically quantified and the correspondence
is by no means perfect.  Typical separations are $\la 1 \Mpc$,
suggesting that the correlation must extend significantly beyond
the virial radius.  Perhaps the clearest example lies along the
line of sight toward PKS 2155--304, which has a pair of
intergalactic \osix absorbers \citep{shull03} offset by $\pm 400
\kms$ from a galaxy group.  Using \emph{Chandra},
\citet{fang02b} also found an \oeight absorber coincident with
the galaxy group, although subsequent observations with
\emph{XMM} failed to confirm this detection.  Because the line
of sight passes near the barycenter of the group, the \oeight
could correspond to virialized gas while the \osix systems probe
gas that has not yet reached the virial radius but has
nevertheless been shocked during infall.  This is somewhat
similar to the interpretation of \citet{shull03}, who argued
that the \osix and \oeight could trace multiphase infalling gas,
although in their case they placed the gas in a single shocked
region.  However, we note that the \oseven absorbers discovered
by \citet{nicastro04} appear to show no correlation with galaxy
systems.  A more detailed study of the relation between galaxies
and the known absorbers should shed light on the physical origin
and significance of these systems.  For example, photoionized
\osix absorbers need not correlate with galaxies because they
are not part of the filamentary structure.

Our model may also apply to the Local Group absorber.
\citet{nicastro02} noted that all three ionization states appear
along the line of sight to PKS 2155--304 with $\Nosix \sim
10^{14} \colden$, $\Noseven \sim 4 \times 10^{15} \colden$, and
$\Noeight \sim 5 \times 10^{15} \colden$.  They modeled the
absorber as a single phase and favored a model with low-density
photoionized gas.  The comparable abundance of \oeight and
\oseven contrasted with the relatively large \osix column ruled
out CIE.  As a result the absorber was forced to have a physical
size $\sim 5 \Mpc$.  In our picture, the transitions sample
different spatial locations and shocks.  We can easily
accommodate $\Noseven/\Nosix$ at infall shocks associated with
halo masses $M \sim 10^{13} M_\odot$.  The column density
$\Noeight$ would be too small in such a shock; however, our
privileged position near the center of the Milky Way and Local
Group implies that this transition could be sampling a second,
nearby shocked component at $T_{\rm vir} \sim 2 \times 10^6
\kel$.  Thus the X-ray absorbers could be cosmological but no
more than a few hundred kpc in extent.  This is reassuring
because, if every dark-matter halo comparable to the Local Group
had a 5 Mpc hot gaseous halo, we would expect to find several
times as many high column-density \oseven and \oeight absorbers
as we see or as appear in simulations \citep{fang02}.  In that case, the
PKS 2155--304 line of sight would have to be unusual; however,
comparable absorption columns appear along many other lines of
sight through the Local Group \citep{rasmussen03}.

Our simple model is clearly not sophisticated enough for
detailed predictions of the absorber statistics.  For those
purposes, cosmological simulations are much more appropriate
\citep{cen01,fang01,fang02}.  \citet{chen03} have presented the
most detailed treatment to date. They showed that modern
simulations do reproduce the observed number densities, at least
to the (admittedly limited) accuracy of the current measurements
and given an {\it ad hoc} (though reasonable) metallicity distribution.  We have taken a complementary approach using analytic arguments to estimate the typical column densities and to parameterize the number densities in terms of the known abundance of collapsed objects.  We believe that a combination of the two approaches should yield a much deeper physical understanding of the substantial baryon reservoir probed by these observations.

We are grateful to R. Cen and J.~P. Ostriker for the use of their
simulation results in Section~\ref{sims} and to C. Danforth and
J.~M. Shull for sharing the data used in Fig.~\ref{fig:o6igm}.  We
thank T. Fang and R. Dav\'{e} for helpful conversations, A. Kravtsov
for enlightening discussions, and M. Sako for clarifying issues
associated with some of the early reports of IGM X-ray-absorption
detections.  SRF thanks the Kavli Institute for Theoretical Physics,
where part of this work was completed under NSF PHY99-07949. This work
was supported at Caltech in part by NASA NAG5-11985 and DoE
DE-FG03-92-ER40701.


\end{document}